\documentclass[aps,prl,twocolumn,superscriptaddress]{revtex4-1}
\usepackage{graphicx}

\begin{document}

\title{Bound States in the Continuum on Periodic Structures: \\
  Perturbation Theory and Robustness} 

\author{Lijun Yuan}
\affiliation{College of Mathematics and Statistics, Chongqing Technology and Business University, Chongqing, 
China}

\author{Ya Yan Lu}
\affiliation{Department of Mathematics, City University of Hong Kong,
  Hong Kong} 

\date{\today}

\begin{abstract}
On periodic structures, a bound state in the continuum (BIC) is 
a standing or propagating Bloch wave with a frequency
in the radiation continuum. Some BICs (e.g., antisymmetric standing
waves) are symmetry-protected, since they have incompatible symmetry
with outgoing waves in the radiation channels. The propagating
BICs do not have this symmetry mismatch, but they still depend
crucially on the symmetry of the 
structure. In this Letter, a perturbation theory is 
developed for propagating BICs on two-dimensional periodic
structures. The study shows that these BICs are robust against
structural perturbations that preserve the symmetry, indicating that 
these BICs are in fact implicitly protected by symmetry.
\end{abstract}

\maketitle


A periodic structure sandwiched between two homogeneous media can be
considered as a (periodic) waveguide. Guided modes on such a
structure usually exist below the light line, so that plane waves
with the same frequency and consistent wavevector can only decay
exponentially in the surrounding homogeneous media. However, it is
known that on periodic structures, guided modes can also exist above
the light line
\cite{bonnet94,padd00,tikh02,shipman03,shipman07,lee12,hu15,
port05,mari08,ndan10,hsu13_1,hsu13_2,bulg14b,bulg15,bulg16,gansch16,gao16,yuan17}, 
and they are special cases of bound states in the 
continuum (BICs), a notion originally introduced by Von 
Neumann and Wigner \cite{neumann29}. Due to their interesting properties and potentially important 
applications, optical BICs have recently attracted much attention 
\cite{hsu16}.  Mathematically, the BICs 
correspond to discrete 
eigenvalues in continuous spectra, and are related to the
nonuniqueness of diffraction problems \cite{bonnet94}. 
BICs are also known to exist on  
waveguides with local distortions  
\cite{urse87,evans91,gold92,evans94,port98,bulg08},  and on waveguides  
with lateral leaky structures \cite{plot11,moli12,weim13,zou15}.  
Related to a BIC on a periodic structure, diffraction problems for
given incident waves exhibit interesting properties such as nonuniqueness, 
total transmission, total reflection, and discontinuities in transmission
(reflection) coefficients \cite{bonnet94,shipman05,shipman12}. These
  properties can be  explored in filtering, sensing and switching
  applications. Near a BIC, there is a family of resonant modes
  with Q-factors approaching infinity. The resulting strong   field
  enhancement  and light confinement can be used to design high-quality laser
  \cite{kante17}, enhance nonlinear optical effects \cite{lijun17pra},
  quantum optical effects, and other emission processes.

On periodic structures, the so-called symmetry-protected BICs
are well-known 
\cite{bonnet94,padd00,tikh02,shipman03,shipman07,lee12,hu15}. They
cannot couple to the outgoing 
waves in the radiation channels due to a symmetry mismatch. The more
interesting BICs are the propagating Bloch modes that do not have 
incompatible symmetry with the outgoing waves
\cite{port05,mari08,ndan10,hsu13_1,hsu13_2,bulg14b,bulg15,bulg16,gansch16,gao16,yuan17}. It
has been observed that the propagating BICs are robust against
structural changes that preserve the relevant symmetry
\cite{hsu13_1,hsu13_2}. If the structure is slightly changed (keeping
the relevant symmetry), the original BIC is simply shifted to a new
one with a slightly different frequency and a slightly different
wavevector.  This phenomenon was investigated numerically for
photonic crystal slabs and periodic arrays of spheres by a number of authors 
\cite{yang14,zhen14,bulg17prl}. Zhen {\it et al.} \cite{zhen14} 
considered the polarization directions of the far-field radiation and
suggested that the Bloch BICs on photonic crystal slabs are topologically
protected. Bulgakov and Maksimov 
\cite{bulg17prl} reached the same conclusion for BICs on a periodic array of
spheres by considering the phase
singularities of a certain coupling coefficient. 
 In this Letter,
we present a perturbation theory for BICs on general periodic structures. Our
result reveals the link between the relevant symmetry and the
robustness of BICs directly and constructively. Namely, if the
periodic structure is changed with a small perturbation that preserve
the symmetry, the BIC is shifted 
to a nearby one which can be solved by the perturbation method. On the
other hand, if the perturbation does not preserve the symmetry, the
method breaks down which indicates that no BICs exist in the
neighborhood. 


We consider a two-dimensional (2D) periodic structure given by a real
dielectric function $\epsilon(x,y)$ which is periodic in $y$ with 
period $L$ and has reflection symmetry in both $x$ and $y$
directions.
The reflection symmetry implies that 
\begin{equation}
  \label{xyeven}
\epsilon(x,y) = \epsilon(-x,y)= \epsilon(x,-y)
\end{equation}
for all $(x,y)$. In addition, we assume that the periodic structure is bounded in the $x$
direction by $|x| < D$ and surrounded by vacuum, i.e., 
$\epsilon(x,y) = 1$ for $|x| > D$. 
For the $E$ polarization, the $z$ component of the
electric field, denoted as $u$, satisfies the following Helmholtz
equation 
\begin{equation}
  \label{helm}
[ \partial_x^2 + \partial_y^2 + k^2 \epsilon(x,y) ] u = 0,
\end{equation}
where $k=\omega/c$ is the free-space wavenumber, $\omega$ is the
angular frequency, and $c$ is the speed of light in vacuum. 

A guided mode on this periodic structure is a special solution
of Eq.~(\ref{helm}) given in the Bloch form 
\begin{equation}
  \label{bloch}
u(x,y) = e^{i \beta y} \phi(x,y), 
\end{equation}
where $\beta$ is the Bloch wavenumber, $\phi$ is periodic in $y$ with
period $L$, and $\phi(x,y) \to 0$ as $|x| \to \infty$. 
The Bloch wavenumber $\beta$ can be restricted to the interval 
$[-\pi/L, \pi/L]$. In terms of
$\phi$, Eq.~(\ref{helm}) becomes ${\cal L} \phi = 0$, where 
\begin{equation}
  \label{defL}
\quad {\cal L} = \partial_x^2 + \partial_y^2 +
2i\beta \partial_y + k^2 \epsilon(x,y) - \beta^2.
\end{equation}
A BIC is a special guided mode above the light line, i.e., $k >
|\beta|$. In the surrounding free space, plane waves with the same
frequency and compatible $y$-dependence have
wavevectors $(\pm \alpha_j, \beta_j)$, where 
\begin{equation}
\beta_j = \beta + 2\pi j/L, \quad \alpha_j = \sqrt{k^2 - \beta_j^2},  
\end{equation}
and $\alpha=\alpha_0$ is positive. We further assume that $k < 2\pi/L
- |\beta|$,  then if $j\ne 0$, $\alpha_j =  i \sqrt{\beta_j^2 - k^2}$ is pure
imaginary. Therefore, for all  $j\ne 0$, the plane waves with
wavevectors $(\pm \alpha_j, \beta_j)$ are evanescent. The only
opening diffraction channel corresponds to plane waves with
wavevectors $(\pm \alpha, \beta)$. 
For $|x| > D$, the BIC can be expanded in plane waves as
\begin{equation}
  \label{bicFour}
  u(x,y) = \sum_{j=-\infty}^\infty c_j^\pm e^{ i (\beta_j y \pm
    \alpha_j x)}, 
\end{equation}
for $x>D$ and $x< -D$, respectively. Since $u \to 0$ as $|x|\to
\infty$, we must have $c_0^\pm = 0$. In addition, it can be assumed that the
BIC is either even in $x$ or odd in $x$. Notice that 
if $u(x,y)$ is a BIC, then so is $u(-x,y)$, we can construct even or
odd BICs from $u(x,y)+u(-x,y)$ or $u(x,y)-u(-x,y)$. It is clear that
$c_j^+ = \pm c_j^-$ depending on whether $u$ is even or odd in $x$. 

It can be easily verified that if $u$ is a BIC, then so it
$\overline{u}(x,-y)$, where $\overline{u}$ denotes the complex 
conjugate of $u$, and they have the same frequency and same Bloch
wavenumber $\beta$. We further assume that $u$ is non-degenerate
(i.e. single), then there must be a constant $C$, such that 
$u(x,y) = C \overline{u}(x,-y)$. 
Let $\Omega$ be one period of the structure given by
\begin{equation}
  \label{Omega}
\Omega = \left\{ (x,y) \ | \ -\infty < x < \infty, \ -L/2<y< L/2 \right\},
\end{equation}
then the integral of $|u|^2$ on $\Omega$
is finite, since $u$ decays to zero exponentially as $|x|$ is
increased. Substituting the above into this integral, we get
$|C|=1$. If $C = e^{ i 2\tau}$ for 
some real number $\tau$, then $w = e^{-i \tau} u$ is also a
BIC, and it satisfies $w(x,y) = \overline{w}(x,-y)$. Therefore,
without loss of generality, we can assume $C=1$, or 
\begin{equation}
  \label{pt}
u(x,y) = \overline{u}(x,-y).  
\end{equation}
The above is a case of the ${\cal PT}$-symmetry. 
It is clear that $\phi$ given in (\ref{bloch}) satisfies the same 
${\cal PT}$-symmetry. 

For the periodic structure, we also consider diffraction problems for incident plane waves with the same frequency and the same wavenumber
($y$-component of the wavevector, 
$\beta$). To obtain a solution $\tilde{v}_e$ that is even
in $x$, we specify two 
incident waves  $\exp[ i (\beta y \pm \alpha  x)]$ for $x<-D$ and
$x>D$, respectively.
The function $\tilde{v}_e$ can be assumed to be even in $x$, since
otherwise, it can be replaced by 
$[\tilde{v}_e(x,y)+ \tilde{v}_e(-x,y)]/2$ which solves the same 
diffraction problem. For large $|x|$, $\tilde{v}_e$ has the following 
asymptotic form 
\[
\tilde{v}_e(x,y) \sim e^{ i (\beta y \pm \alpha  x)} + S_e  e^{ i
  (\beta y \mp \alpha x)}, \quad x \to \mp \infty,
\]
where $S_e=e^{2i \theta_e}$ is a complex number satisfying $|S_e|=1$. Let $v_e =
e^{ -i \theta_e} \tilde{v}_e$, then we can assume that $v_e$ satisfies
the ${\cal PT}$-symmetry, 
since otherwise, it can be replaced by
$[v_e(x,y)+\overline{v}_e(x,-y)]/2$ which solves the same diffraction
problem with the same asymptotic behavior as infinity. Similarly, we can construct a diffraction solution $v_o(x,y)$
which is odd in $x$ and is ${\cal PT}$-symmetric. Finally, we 
define $\varphi_e$ and $\varphi_o$ as in Eq.~(\ref{bloch}). That is, 
\begin{equation}
v_e(x,y) = e^{i \beta y} \varphi_e(x,y), \ 
v_o(x,y) = e^{ i \beta y} \varphi_o(x,y),  
\end{equation}
where $\varphi_e$ and $\varphi_o$ are periodic in $y$ with period
$L$, are ${\cal PT}$-symmetric, and are even and odd in $x$,
respectively.  


Assuming that the periodic structure given by $\epsilon(x,y)$ has a
non-degenerate BIC $u=e^{i \beta y} \phi$ for free-space wavenumber $k$
(frequency $\omega$) and Bloch wavenumber $\beta$, we now
consider a perturbed structure given by 
\begin{equation}
\tilde{\epsilon}(x,y) = \epsilon(x,y)+ \delta F(x,y),
\end{equation}
where $\delta$ is a small real number and $F$ is a real $O(1)$ function
satisfying $F(x,y)=0$ for $|x|>D$. We look for a BIC $\tilde{u} = e^{
  i \tilde{\beta} y} \tilde{\phi}$ with 
Bloch wavenumber $\tilde{\beta}$ and free-space wavenumber
$\tilde{k}$. In the perturbation method, $\tilde{\phi}$, $\tilde{k}$
and $\tilde{\beta}$ are expanded in power series of $\delta$:
\begin{eqnarray}
  \label{expphi} && \tilde{\phi} = \phi + \delta \phi_1 + \delta^2 
                    \phi_2 + \cdots \\
  \label{expbeta} && \tilde{\beta} = \beta + \delta \beta_1 + \delta^2 
                    \beta_2 + \cdots \\
  \label{expbeta} && \tilde{k} = k + \delta k_1 + \delta^2 k_2 + \cdots
\end{eqnarray}
Inserting the above expansions into the equation for $\tilde{\phi}$,
i.e., $\tilde{\cal L}
\tilde{\phi} = 0$, where $\tilde{\cal L}$ is defined as ${\cal L}$ given in
(\ref{defL}), with $\epsilon$, $\beta$ and $k$ replaced by
$\tilde{\epsilon}$, $\tilde{\beta}$ and $\tilde{k}$, respectively, we
obtain a sequence of equations for $\phi_1$, $\phi_2$, etc. They can
be written as 
\begin{equation}
  \label{eq4j}
{\cal L} \phi_j =  B_1 \beta_j + B_2 k_j - C_j, 
\end{equation}
for any integer $j \ge 1$, where $B_1$ and $B_2$ are independent of
$j$, and $C_j$ does not involve $\beta_j$ and $k_j$. More precisely,
\begin{eqnarray}
\label{B1} B_1 &=& 2 \beta \phi - 2 i \partial_y \phi, \\
\label{B2} B_2 &=&  -2 k \epsilon \phi, \\
\label{C1} C_1 &=& k^2 F \phi, \\
\label{C2} C_2 &=& (k_1^2 \epsilon - \beta_1^2 + 2 k k_1 F) \phi \cr
     & +& (2 kk_1 \epsilon - 2 \beta \beta_1 + k^2 F) \phi_1 
 + 2 i    \beta_1 \partial_y \phi_1, 
\end{eqnarray}
and the general formula for $C_j$ is
\begin{eqnarray}
&& C_{j} = \left[\sum_{l=1}^{j-1} (k_l k_{j-l} \epsilon - \beta_l \beta_{j-l})
+ F \sum_{l=0}^{j-1} k_l k_{j-1-l} \right] \phi \cr
&& + \sum_{n=1}^{j-1} \left[ \sum_{l=0}^{n} \left( k_l k_{n-l}
   \epsilon - \beta_l \beta_{n-l} \right) + F \sum_{l=0}^{n-1} k_l 
   k_{n-1-l} \right] \phi_{j-n} \cr
\label{Cj} && + 2 i \sum_{n=1}^{j-1} \beta_n \partial_y \phi_{j-n},
\end{eqnarray}
where $\beta_0=\beta$ and $k_0=k$. 

The perturbation theory is developed to investigate the existence of
BICs on the perturbed structure. If there is a BIC 
 on the perturbed structure $\tilde{\epsilon}$ and it is 
near the original one, we expect to carry out the perturbation process
successfully, that is,  $\beta_j$, $k_j$ and $\phi_j$ can be solved,
$\beta_j$ and $k_j$ are real, and $\phi_j$ decays to zero exponentially
 as $|x| \to \infty$. On the other hand, if the perturbation process
 breaks down, that is, $\beta_j$ and/or $k_j$ become complex, or $\phi_j$
 does not decay to zero at infinity, then we believe there is no BIC
 near the original one.

In the $j$th step, $\beta_j$ and $k_j$ are solved from the 
linear system
\begin{equation}
  \label{2by2}
{\bf A} \left[ \matrix{ \beta_j \cr k_j} \right] 
= \left[ \matrix{ a_{11} & a_{12} \cr a_{21} & a_{22} } \right] 
\left[ \matrix{ \beta_j \cr k_j} \right] 
= \left[ \matrix{ b_{1j} \cr b_{2j} } \right], 
\end{equation}
where ${\bf A}$ is a $2\times 2$ matrix, 
\[
a_{11} = \int_\Omega \overline{\phi} B_1 d{\bf r}, \
a_{12} = \int_\Omega \overline{\phi} B_2 d{\bf r}, \
b_{1j} = \int_\Omega \overline{\phi} C_j d{\bf r}, 
\]
$B_1$, $B_2$ and $C_j$ are given in Eqs.~(\ref{B1}), (\ref{B2}) and
(\ref{Cj}), respectively, and ${\bf r}=(x,y)$. The second row depends
on whether $\phi$ is even or odd in $x$. If $\phi$ is even in
$x$, then 
\[
a_{21} = \int_\Omega \overline{\varphi}_e B_1 d{\bf r}, \
a_{22} = \int_\Omega \overline{\varphi}_e B_2 d{\bf r}, \
b_{2j} = \int_\Omega \overline{\varphi}_e C_j d{\bf r}.
\]
If $\phi$ is odd in $x$, then we replace $\varphi_e$ by $\varphi_o$. 
The first equation in (\ref{2by2}) is obtained by 
multiplying  Eq.~(\ref{eq4j}) by $\overline{\phi}$ 
and integrating
both sides on $\Omega$. It can be verified that 
$\int_\Omega \overline{\phi} {\cal L} \phi_j d{\bf r} = 0$, and 
$a_{11}$ and $a_{12}$ are both real. The second equation in
(\ref{2by2}), for the case when $\phi$ is even in $x$, is a result of requiring 
\begin{equation}
  \label{evencon}
\int_\Omega \overline{\varphi}_e ( B_1 \beta_j + B_2 k_j - C_j) d{\bf 
  r} = 0.  
\end{equation}
Since $\varphi_e$ is ${\cal PT}$-symmetric, it can be shown that $a_{21}$ and
$a_{22}$ are also real.  If matrix ${\bf A}$ is invertible, 
$\beta_j$ and $k_j$ can be determined, then $\phi_j$ can be solved from
Eq.~(\ref{eq4j}). The right hand side of Eq.~(\ref{eq4j})
represents a given  source distribution.  

In the following, we show that if ${\bf A}$ is invertible and 
if $F$ has the same symmetry as $\epsilon$, that is, 
\begin{equation}
\label{Fsym}
  F(x,y)=F(x,-y)=F(-x,y)  
\end{equation}
for all $(x,y)$, then the above perturbation process can be carried
out successfully. 
For $j=1$, since both $\phi$ and $\varphi_e$ (or
$\varphi_o$) are ${\cal PT}$-symmetric, it is easy to see that
$b_{11}$ and $b_{21}$ are real. Therefore, $\beta_1$ and $k_1$ can be uniquely solved and they are real. The
governing equation for $\phi_1$ has a nonzero source term in the right hand
side, thus in general, $\phi_1$ should satisfy outgoing radiation
conditions as $x \to \pm \infty$. When $\phi$ is even in $x$, the right
hand side of Eq.~(\ref{eq4j}) (for $j=1$) is even in $x$, thus, we can
assume $\phi_1$ is even in $x$, since otherwise,  it can be replaced
by $[ \phi_1(x,y) + \phi_1(-x,y)]/2$. Since $\phi$ is known to decay
exponentially as $|x| \to \infty$, it is easy to construct particular
solutions for Eq.~(\ref{eq4j}), $j=1$, for $|x| > D$, and these
particular solutions decay exponentially as $|x| \to \infty$. The
general solution for $\phi_1$ is a sum of the particular solution and
a solution of the homogeneous equation. Because of the symmetry in $x$
and the outgoing radiation condition, we have 
\begin{equation}
\phi_1(x,y) \sim d_1 e^{ \pm i \alpha x}, \quad x \to \pm \infty,
\end{equation}
for an unknown coefficient $d_1$. To show that $d_1$ is actually zero,
we choose $h>D$, multiply Eq.~(\ref{eq4j}) for $j=1$, by
$\overline{\varphi}_e$, and integrate on the rectangular domain
$\Omega_h$ given by $|x| < h$ and $|y| < L/2$. From Eq.~(\ref{evencon}), it is clear that 
\[
\lim_{h\to \infty} \int_{\Omega_h} \overline{\varphi}_e  {\cal L}
\phi_1 d{\bf r} = 0.
\]
Using integration by parts and the asymptotic formulas for $\varphi_e$
and $\phi_1$, it can be shown that the left hand side above is $4i d_1
\alpha L e^{-i  \theta_e}$. Therefore, we must have $d_1=0$. Finally,
since the right hand side of Eq.~(\ref{eq4j}) is ${\cal
  PT}$-symmetric, we can assume $\phi_1$ is also ${\cal
  PT}$-symmetric, since otherwise, we can replace $\phi_1$ by 
$[\phi_1(x,y) + \overline{\phi}_1(x,-y)]/2$. 

The same reasoning can be used in all perturbation steps for $j \ge
2$.  If $F$ satisfies Eq.~(\ref{Fsym}) and ${\bf A}$
is invertible, in the $j$th step,  we can show that $\beta_j$ and $k_j$ are real, $\phi_j 
\to 0$ exponentially as $x \to \pm \infty$, $\phi_j$ is even or odd in 
$x$ (same as $\phi$), and $\phi_j$ is ${\cal PT}$-symmetric. 

If $F$ does not satisfy Eq.~(\ref{Fsym}), then the perturbation
process is likely to fail. In the first step, if $\phi$ is even in
$x$, in order to have a real $b_{21}$, we need 
$\int F \phi \overline{\varphi}_e d{\bf r}$ to be real. This implies 
\begin{equation}
  \label{nece1}
\int_\Omega [ F(x,y)-F(x,-y)] \phi \overline{\varphi}_e d{\bf r} = 0. 
\end{equation}
Meanwhile, Eq.~(\ref{evencon}) should remain valid when $\varphi_e$ is
replaced by $\varphi_o$. This gives rise to the condition $\int F
\phi \overline{\varphi}_o d{\bf r} = 0$, which can be written as 
\begin{equation}
  \label{nece2}
\int_\Omega [ F(x,y)-F(-x,y)] \phi \overline{\varphi}_o d{\bf r} = 0. 
\end{equation}
If $\phi$ is odd in $x$, we swap $\varphi_e$ and $\varphi_o$ in Eqs.~(\ref{nece1}) and
(\ref{nece2}).  If $F$ does not satisfy Eq.~(\ref{nece1}) or
(\ref{nece2}), then the perturbation process fails in the first
step. If $F$ satisfies these two conditions, it may be possible to
find $\phi_1$ that decays to zero as $|x| \to \infty$, but the
perturbation method can still fail in the second step. In order to obtain real
$\beta_2$ and $k_2$, $F$ must satisfy additional
conditions that involve $\phi_1$. It is clear that in order to to carry out the
perturbation process to all steps, $F$ must satisfy an infinite sequence of
conditions. Therefore, if $F$ does not satisfy
Eq.~(\ref{Fsym}), the perturbation process is likely to fail. When
this happens, we believe there is no BIC near the original one of the
unperturbed structure. 

The perturbation process also requires a non-singular matrix ${\bf A}$.
This condition is independent of the
perturbation profile $F$, since ${\bf A}$ involves only the BIC $\phi$ and the diffraction
solutions $\varphi_e$ or $\varphi_o$.  When the structure is
continuously varied while the required symmetry is preserved, a BIC may
cease to exist. This happens when the frequency and wavenumber of the
BIC have reached the condition for the opening of the second 
diffraction channel, or when the the BIC becomes a
symmetric standing wave which is even in $y$. 

\begin{figure}[htb]
\centering 
\includegraphics[scale=0.5]{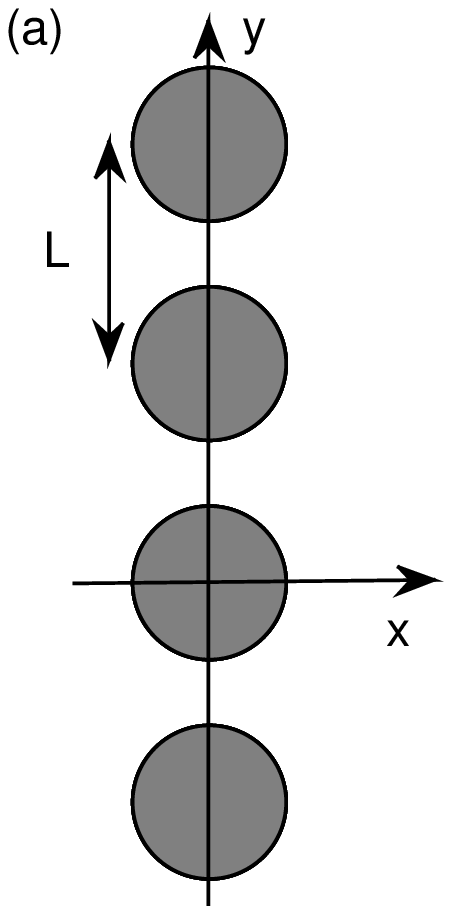} 
\hspace{0.4cm}
\includegraphics[scale=0.5]{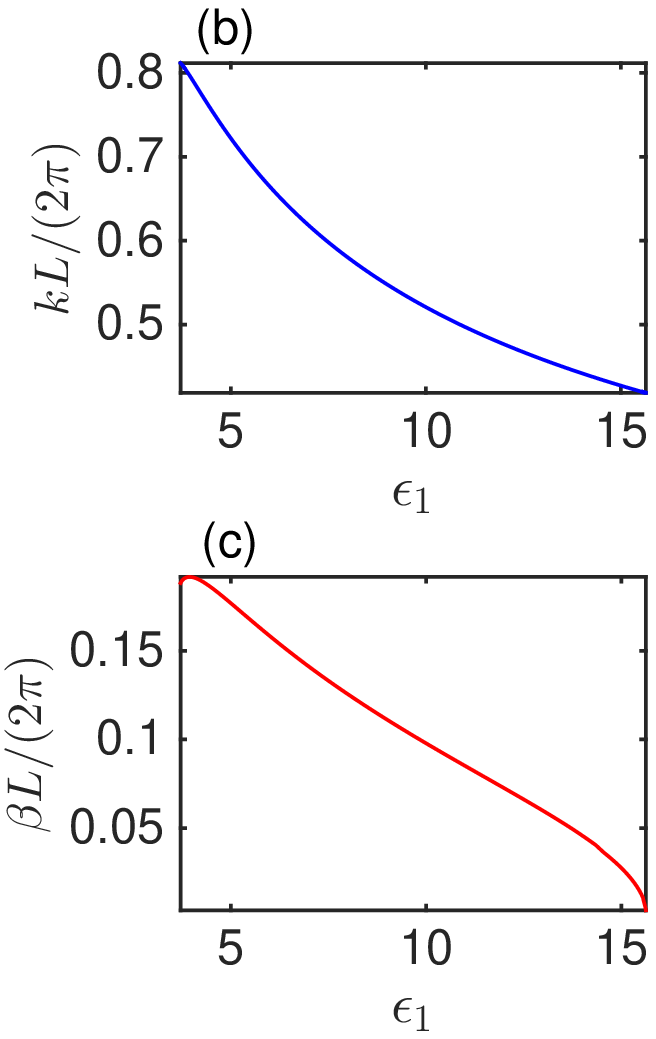}
\caption{(a) A periodic array of circular dielectric rods with radius
  $0.35L$ and dielectric constant $\epsilon_1$. (b)
  Free-space wavenumber $k$ as a function of $\epsilon_1$ for a family of
  $x$-even BICs on the periodic array. (c) Bloch wavenumber $\beta$ as
  a function of $\epsilon_1$ for the same BIC family.}
\label{fig1}
\end{figure}

\begin{figure}
\centering 
\includegraphics[scale=0.5]{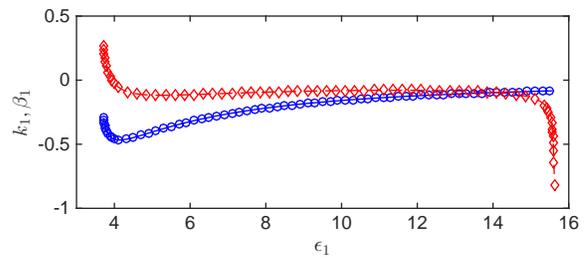}
\caption{First order perturbation results $k_1$ (shown as ``$\circ$'')
  and $\beta_1$ (shown as ``$\diamond$'') of a BIC family 
  for different values of $\epsilon_1$, and finite difference approximations of
  $dk/d\epsilon_1$ (solid blue line) and $d\beta/d\epsilon_1$ (red
  dashed line), all in unit $1/L$.}
\label{fig2}
\end{figure}

To validate the perturbation result, we consider a periodic array of
dielectric rods on which various BICs exist \cite{hu15,bulg14b,yuan17}.  
In particular, if the radius of the rods is $0.35L$,  there is
a family of $x$-even BICs for $3.711 < \epsilon_1 \le 15.701$, where
$\epsilon_1$ is the dielectric constant of the rods. The free-space
wavenumber $k$ and Bloch wavenumber $\beta$ of the BICs are shown as functions of
$\epsilon_1$ in Figs.~\ref{fig1}(b) and \ref{fig1}(c). 
Notice that for $\epsilon_1 = 15.701$, the BIC becomes a symmetric
standing wave (even in $y$). The perturbation theory is applicable, if
we let $F(x,y)=1$ for $(x,y)$ in the rods and $F(x,y)=0$ otherwise,
and in that case, $\beta_1$ and $k_1$ in the first order perturbation
are simply the derivatives of $\beta$ and $k$ with respect to
$\epsilon_1$. In Fig.~\ref{fig2}, 
we show the perturbation results $k_1$ and $\beta_1$ for different
values of $\epsilon_1$, and compare them with finite difference
approximations for $dk/d\epsilon_1$ and $d\beta /d\epsilon_1$. 
The excellent agreement in Fig.~\ref{fig2} indicates that our
perturbation theory is correct. As
$\epsilon_1$ tends to $15.701$, $\beta_1$ or $d\beta/d\epsilon_1$
becomes negative infinity. In fact, ${\bf A}$ is singular for
$\epsilon_1=15.701$. As $\epsilon_1 \to 3.711$, the BIC ceases to exist
because $k$ and $\beta$ approach the condition $k = 2\pi/L - \beta$, that
corresponds to the opening of the second diffraction channel. 
In this limit, the entries of ${\bf A}$ tend to infinity, but $\beta_1$ and $k_1$
approach constants. 

In summary, a perturbation theory is developed  for BICs on general 2D periodic
structures with reflection symmetry in both $x$ and $y$
directions, and it shows that the propagating BICs are robust with respect
to structural perturbations that preserve the reflection
symmetry. 
The importance of symmetry for propagating BICs on periodic
structures was first realized  by Hsu {\it et al.} \cite{hsu13_1,hsu13_2},
and the existence and robustness of these BICs have been investigated
numerically for particular periodic structures
involving a few parameters \cite{zhen14,bulg17prl}. Our perturbation
theory is analytic, and it is applicable to general 2D periodic
structures and general perturbations. Our perturbation results could also be useful 
for sensitivity analysis and optimal design of periodic
structures. In this work, we have concentrated on the 2D case for
simplicity. Clearly, it is worthwhile to extend the perturbation
theory to three-dimensional (3D) rotationally symmetric structures with one
periodic direction, or 3D structures with two periodic directions. 

This work was supported by the Basic and Advanced Research Project of
CQ CSTC (cstc2016jcyjA0491), the Science and Technology Research
Program of Chongqing Municipal Education Commission (KJ1706155), the
Program for University Innovation Team of Chongqing (CXTDX201601026),
and City University of Hong Kong (7004669).

\end{document}